\documentclass[journal=jpclcd,manuscript=letter,layout=twocolumn,toc]{achemso}
\usepackage{graphicx}
\usepackage{dcolumn}
\usepackage{bm}
\usepackage{ulem}
\usepackage{xcolor}
\usepackage{color,soul}
\usepackage{amsmath}
\usepackage{hyperref}
\hypersetup{colorlinks,allcolors=blue}
\DeclareUnicodeCharacter{0308}{\"{u}}

\usepackage[version=4]{mhchem}

\title{
Reactive calcium carbonate precipitation from an atomic cluster expansion potential and enhanced sampling
}

\author{Eslam Ibrahim}
\email{eslam.saadibrahim@rub.de}
\affiliation{ICAMS, Ruhr Universit\"at Bochum, 44780 Bochum, Germany}
\author{Yury Lysogorskiy}
\author{Ralf Drautz}
\email{ralf.drautz@rub.de}
\affiliation{ICAMS, Ruhr Universit\"at Bochum, 44780 Bochum, Germany}
\author{Pablo M. Piaggi}
\email{pm.piaggi@nanogune.eu}
\affiliation{CIC nanoGUNE, Tolosa Hiribidea 76, Donostia 20018, San Sebastian, Spain}
\alsoaffiliation{Ikerbasque, Basque Foundation for Science, Bilbao 48013, Spain}

\begin{document}

\begin{abstract}
Calcium carbonate formation from aqueous solution is central to biomineralization and to carbon sequestration through mineral carbonation.
At near-neutral pH, the process is highly reactive, with proton transfer mediating the interconversion between carbonate species.
Most atomistic simulations to date either treat carbonate speciation as fixed or consider proton transfer only in small clusters.
Here, we combine an ab initio–trained atomic cluster expansion (ACE) machine-learning potential for molecular dynamics with enhanced sampling to enable reactive simulations of the early stages of calcium carbonate precipitation at previously inaccessible length and time scales.
We study proton transfer and carbonate speciation in ion pairs and triplets, as well as in the collective aggregation of many ions.
Our simulations with few ions show that ion association provides a favorable pathway for proton transfer, facilitating interconversion between carbonate, bicarbonate, and carbonic acid.
In many-ion systems, proton transfer occurs spontaneously alongside aggregation, and we observe significant changes in the coordination environments as species evolve during the simulations. 
These results show that ion aggregation and chemical reactivity can be strongly coupled during the early stages of nucleation from solution.

\begin{tocentry}
  \includegraphics[width=1\textwidth]{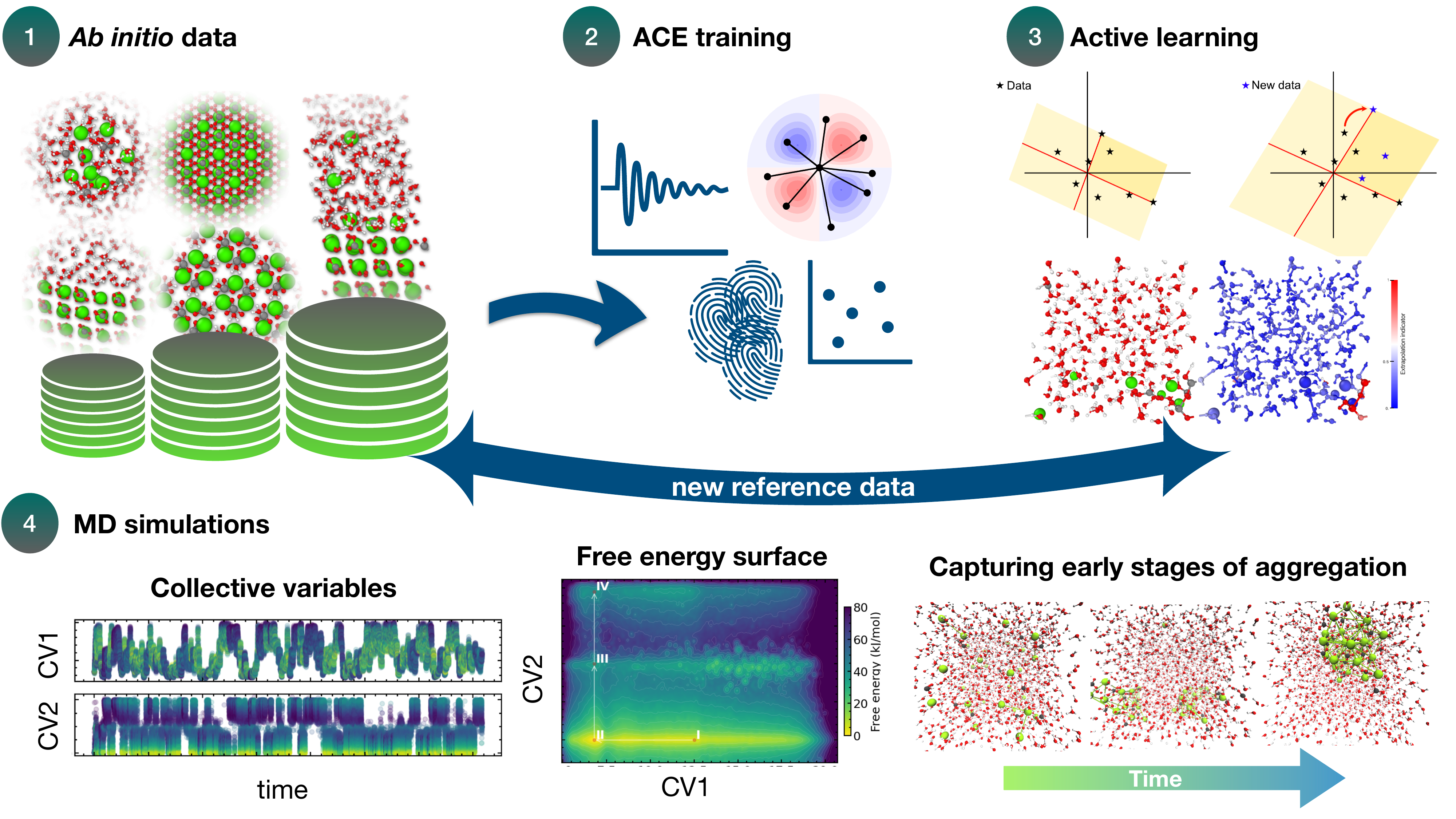}
\end{tocentry}
\end{abstract}

\maketitle

Calcium carbonate (\ce{CaCO3}) formation from aqueous solution underlies biomineralization, ocean chemistry, geological carbon cycling, and carbon sequestration technologies.
The crystallization of \ce{CaCO3} is also technologically consequential, as it leads to scale formation in industrial systems such as boilers and desalination plants.
At the molecular level, these processes are governed not only by ion association but also by chemical reactivity in solution, including proton transfer and speciation dynamics, which directly influence the stability and evolution of early aggregates~\cite{agmon2016protons,de2015crystallization}.
Despite decades of investigation, the molecular mechanisms governing its earliest stages remain incompletely resolved~\cite{di2009theoretical,raiteri2010water,weiner2011crystallization,gebauer2014pre,smeets2017classical}.

Classical descriptions of nucleation portray \ce{CaCO3} formation as the progressive association of stable species into larger clusters~\cite{kashchiev2000nucleation}.
However, this picture remains under active debate, and non-classical pathways involving the progressive assembly of dynamic oligomers~\cite{gebauer2008stable,demichelis2011stable,gebauer2014pre,kimura2022possible} and a  liquid–liquid phase transition \cite{wallace2013microscopic,henzler2018supersaturated} have been proposed.
At near-neutral pH, solutions are predominantly composed of bicarbonate ions (\ce{HCO3-}), whereas at higher pH carbonate  (\ce{CO3^2-}) becomes dominant~\cite{huang2021uncovering}.
While at high pH the formation of \ce{CaCO3} can proceed directly from \ce{CO3^2-}, i.e. without chemical reactivity, at near-neutral pH the phenomenon is mediated by the transformation of \ce{HCO3-} to \ce{CO3^2-} accompanied by the release of \ce{CO2}~\cite{jin2025formation}.
A key unresolved problem is understanding how chemical speciation and ion aggregation are interconnected.
Resolving this issue is central to understanding the microscopic mechanism of early \ce{CaCO3} formation.

Addressing this question requires a framework capable of combining electronic-structure-level chemical accuracy with sampling over nanosecond timescales.
Ab initio molecular dynamics~\cite{car1985unified} based on density-functional theory (DFT) calculations\cite{kohn-1965} provides reliable energetics but is limited in system size and sampling.
Conversely, empirical force fields enable large-scale simulations yet often lack the accuracy necessary to describe chemical equilibria in solution and are usually unable to describe chemical reactivity, i.e., bond forming and breaking~\cite{tribello2009molecular,raiteri2010derivation,raiteri2015thermodynamically}.
The development of machine-learning potentials over the last decade has made it possible to perform reactive simulations at length and time scales beyond those routinely accessible to ab initio molecular dynamics~\cite{behler2007generalized,drautz2019atomic,piaggi2025ab}.
Recently, one of us trained one such potential~\cite{piaggi2025ab} able to describe \ce{CaCO3} in solution and in the solid phase using the SCAN~\cite{sun2015strongly} DFT exchange and correlation functional.
However, limited data for multiple ions in solution hindered the applicability of that model to understand the initial stages of \ce{CaCO3} precipitation.

Here, we build directly on this foundation and develop an atomic cluster expansion (ACE) potential trained on data computed using the revPBE-D3 DFT functional, specifically aimed at describing accurately the intense reactivity during the initial stages of \ce{CaCO3} precipitation.
We combine this ACE potential with enhanced sampling simulations and large-scale molecular dynamics to understand the interplay between ion aggregation and chemical
reactivity during the early stages of precipitation from solution.

\begin{figure*}[hbt!]
    \centering
    \includegraphics[width=\textwidth]{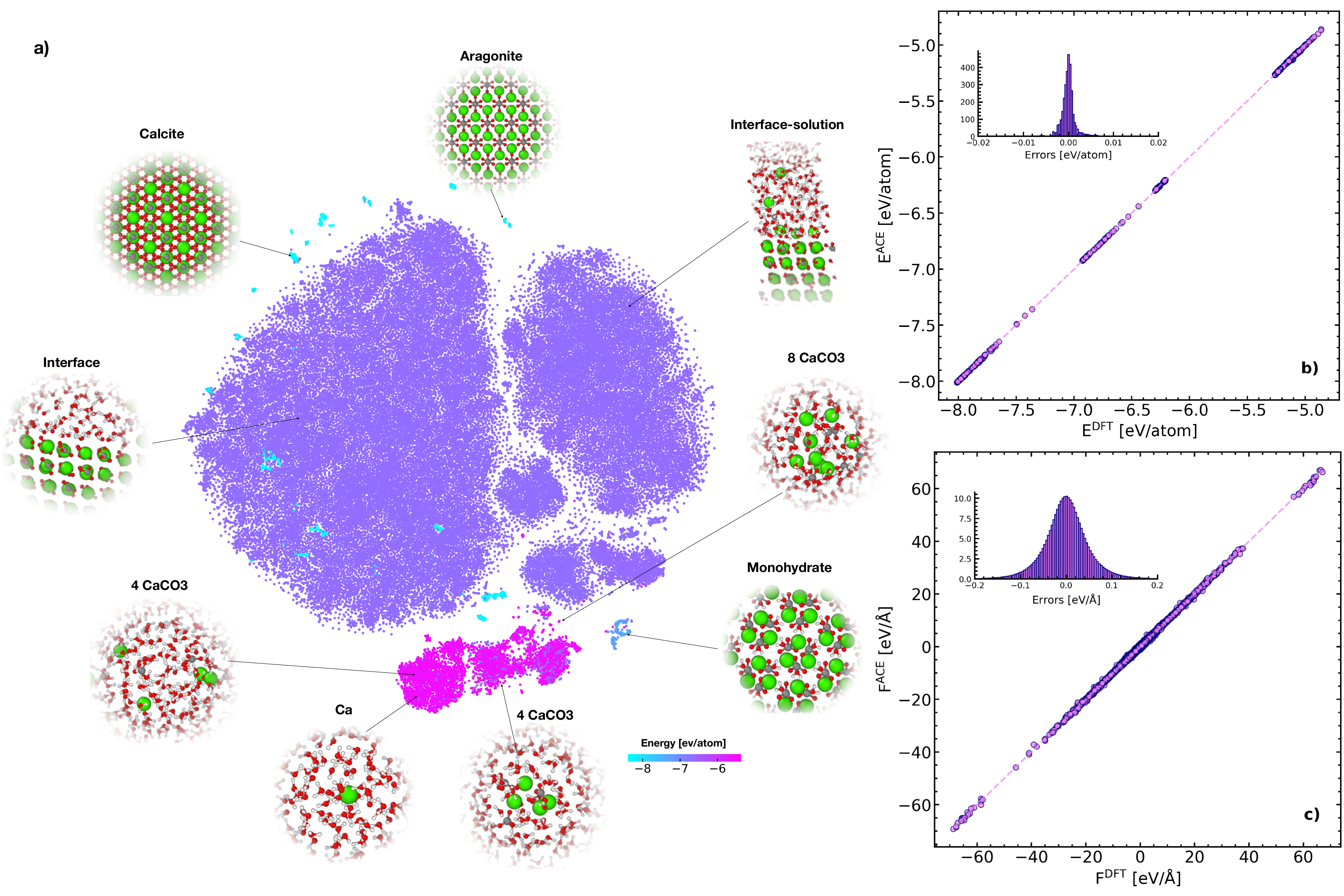}
    \caption{
    (a) Low-dimensional embedding of the ACE training dataset using t-SNE, colored by atomic energy.
    Each point represents a local atomic environment sampled across aqueous solution, ion pairs, multicarbonate clusters, interfacial configurations, and crystalline phases (calcite and aragonite).
    (b) Parity plot comparing ACE-predicted and DFT reference energies, with inset showing the distribution of energy errors of the training dataset.
    (c) Parity plot for atomic forces with corresponding error distribution.
    }
    \label{fig:tsns_caco3}
\end{figure*}


To tackle this problem, we first created a dataset of atomic configurations able to represent the chemically diverse aqueous CaCO$_3$ system across a broad range of coordination environments and protonation states.
Initial configurations were constructed starting from the calcium carbonate dataset reported in our previous work~\cite{piaggi2025ab}, for which all structures were recomputed at the revPBE-D3 level to ensure a consistent electronic-structure reference.
The choice of revPBE-D3 is motivated by our recent development of a general and transferable ACE potential for water~\cite{ibrahim2024efficient,ibrahim2026pd}, which accurately reproduces liquid structure, ice polymorph energetics, and the water phase diagram across a wide thermodynamic range.

We then generated additional configurations through active learning based on running molecular dynamics with enhanced-sampling.
We used aqueous solutions of approximately 600 atoms covering Ca concentrations between 0.5 M to 2.2 M, which corresponds to 2 to 8 Ca atoms.
We considered carbonate-only, bicarbonate-only, and mixed carbonate--bicarbonate compositions, spanning conditions characteristic of carbonate-rich and bicarbonate-rich solutions. The pH was not imposed or calculated explicitly.
To thoroughly sample ion aggregation and reactivity, we performed biased simulations using On-the-fly Probability Enhanced Sampling (OPES)\cite{invernizzi2020rethinking,invernizzi2020unified}, which is an evolution of the Metadynamics method\cite{laio2002escaping}.
We used as collective variables the mean Ca-C coordination number, to foster aggregation, and the number of carbonate species, to promote reactivity.
These CVs were defined in a continuous and differentiable fashion as described in the Methods section.
From the trajectories generated using the biased OPES simulations, we selected configurations for inclusion in the training dataset when the ACE extrapolation grade exceeded a threshold,
identifying environments insufficiently represented in the current model.
Selection followed a D-optimality criterion~\cite{podryabinkin2017active,lysogorskiy2023active}.

In total, 25,819\, configurations comprising 17,208,893\, atomic environments were included in our training dataset. 
These configurations encompass dilute solvated ions, solvent-separated and contact ion pairs, proton-transfer intermediates between carbonate, bicarbonate, and carbonic acid species, hydrated multicarbonate clusters, crystalline polymorphs including calcite and aragonite, and surface configurations.
The structural diversity of the final dataset is illustrated in Fig.~\ref{fig:tsns_caco3}a.
This figure  maps local atomic environments, represented using the ACE descriptors, into a two-dimensional representation using t-distributed stochastic neighbor embedding (t-SNE). Crystalline and disordered regions are clearly separated, while continuous manifolds connect solvated ions to aggregated and interfacial structures, confirming that the dataset spans the relevant regions of configuration space. 

Afterwards, we proceeded to train a model for the interatomic interactions using ACE~\cite{drautz2019atomic}, a formally complete descriptor of the local atomic environments with a systematically convergent body-ordered basis representation of local atomic environments for single and multi-component materials. 
The ACE method enables compact representation of many-body interactions while preserving transferability between aqueous and crystalline phases.
The fitting accuracy for energies and forces is shown in Fig.~\ref{fig:tsns_caco3}b,c.
The final model achieves an energy RMSE of 1.59\,meV\,atom$^{-1}$ and mean absolute error (MAE) of 0.97\,meV\,atom$^{-1}$, together with a force RMSE of 51.4\,meV\,\AA$^{-1}$ and MAE of 37.1\,meV\,\AA$^{-1}$ on the training dataset.
Together, the broad structural coverage and low fitting errors show that the ACE potential reproduces the DFT reference data across the reactive calcium carbonate environments included in the fit.

\begin{figure}[hbt!]
    \centering
    \includegraphics[width=0.5\textwidth]{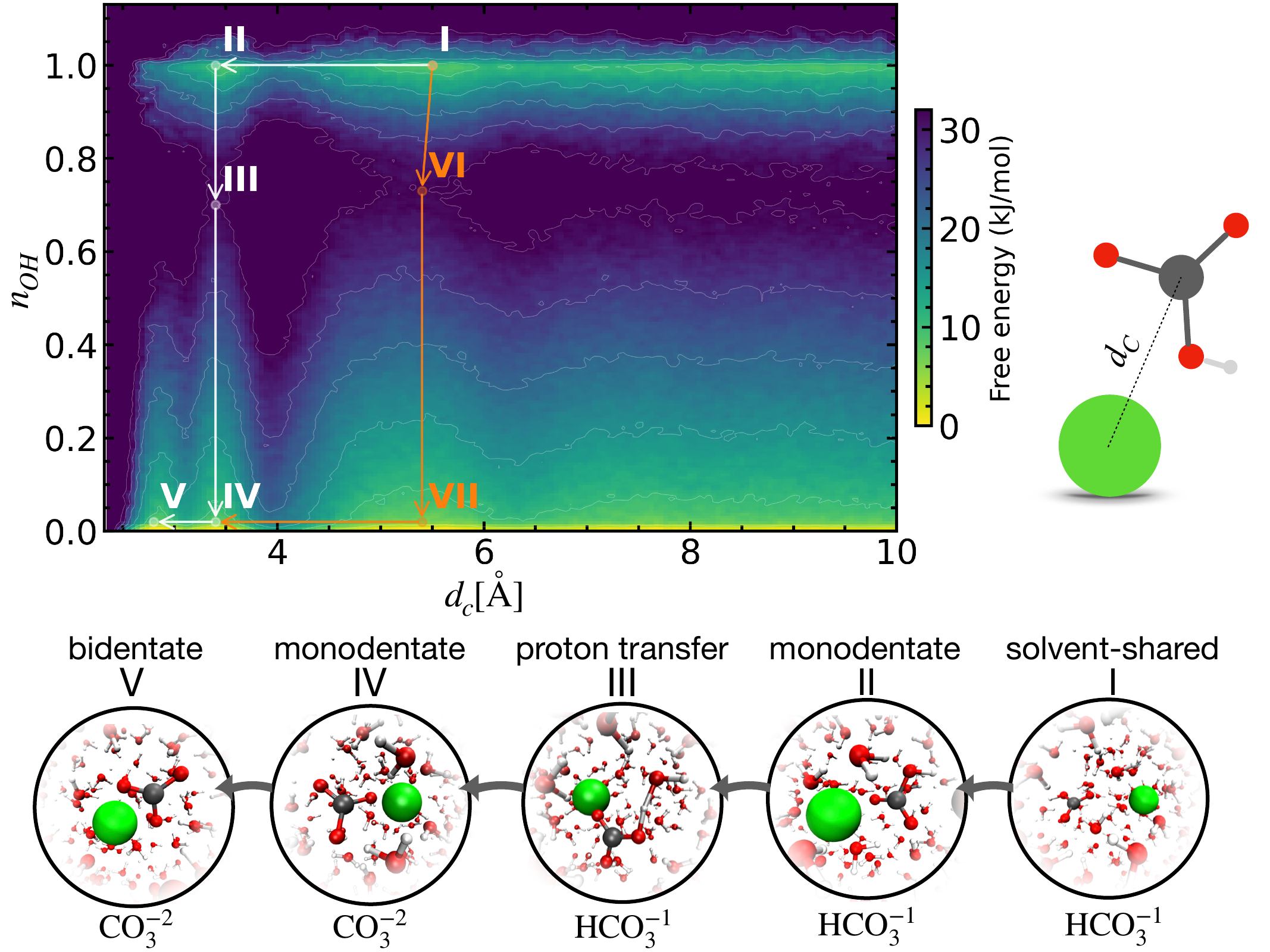}
    \caption{
Free energy landscape of Ca$^{2+}$ and carbonate association in aqueous solution.
The free energy $F(d_{\mathrm{C}}, n_{\mathrm{OH}})$ (color scale) is shown as a function of the Ca–C distance $d_{\mathrm{C}}$ and the carbonate protonation coordinate $n_{\mathrm{OH}}$ (smooth O–H coordination, with $n_{\mathrm{OH}}\!\approx\!0$ for CO$_3^{2-}$ and $n_{\mathrm{OH}}\!\approx\!1$ for HCO$_3^{-}$).
Relevant states are marked with Roman numerals and representative configurations are shown below.
White and orange arrows indicate two possible pathways for ion pairing.
}
    \label{fig:ca_carb}
\end{figure}

We now discuss the results of our simulations.
We begin with the minimal reactive unit: a single Ca$^{2+}$ ion interacting with one carbonate species in water.
We carried out enhanced-sampling OPES simulations using two collective variables: a structural coordinate, the Ca--C separation distance ($d_{\mathrm{C}}$), and a chemical coordinate describing protonation, defined as the coordination between carbonate oxygens and hydrogen atoms (O$_{\mathrm{C}}$--H coordination).
Additional details are provided in the Methods section.
Figure~\ref{fig:ca_carb} reports the two-dimensional free-energy surface as a function of $d_{\mathrm{C}}$ and O$_{\mathrm{C}}$--H coordination.
The landscape contains well-defined basins at coordination values of 0 and 1, corresponding to carbonate and bicarbonate species, respectively.
For each speciation state, we observe solvent-shared configurations (I and VII) and contact ion pairs (II, IV, and V), as well as transition regions associated with proton transfer (III).
Contact ion pairs for carbonate can be further classified into monodentate (IV) and bidentate (V) coordination motifs.
These states are consistent with previous ab initio and machine-learning simulations of aqueous calcium carbonate~\cite{tommaso2008onset,huang2021uncovering,li2024ion,piaggi2025ab}.
Along the minimum-free-energy pathway (white), ion association proceeds from solvated ions to contact ion pair motifs.
Importantly, the pathway for bicarbonate to carbonate transformation goes through a proton-transfer state (III) which becomes accessible only after the formation of the contact ion pair configurations, indicating that proton transfer is enabled by structural association rather than occurring independently in solution.
An alternative pathway through the solvent-shared region is also observed at somewhat larger ion separations, as shown in orange in Figure~\ref{fig:ca_carb}.
These findings establish that changes in chemical speciation are favored by the coordination environment created during ion association and are consistent with previous theoretical and experimental studies~\cite{wu2024bicarbonate,piaggi2025ab,zhu2025molecular}.

\begin{figure}[hbt!]
    \centering
    \includegraphics[width=0.5\textwidth]{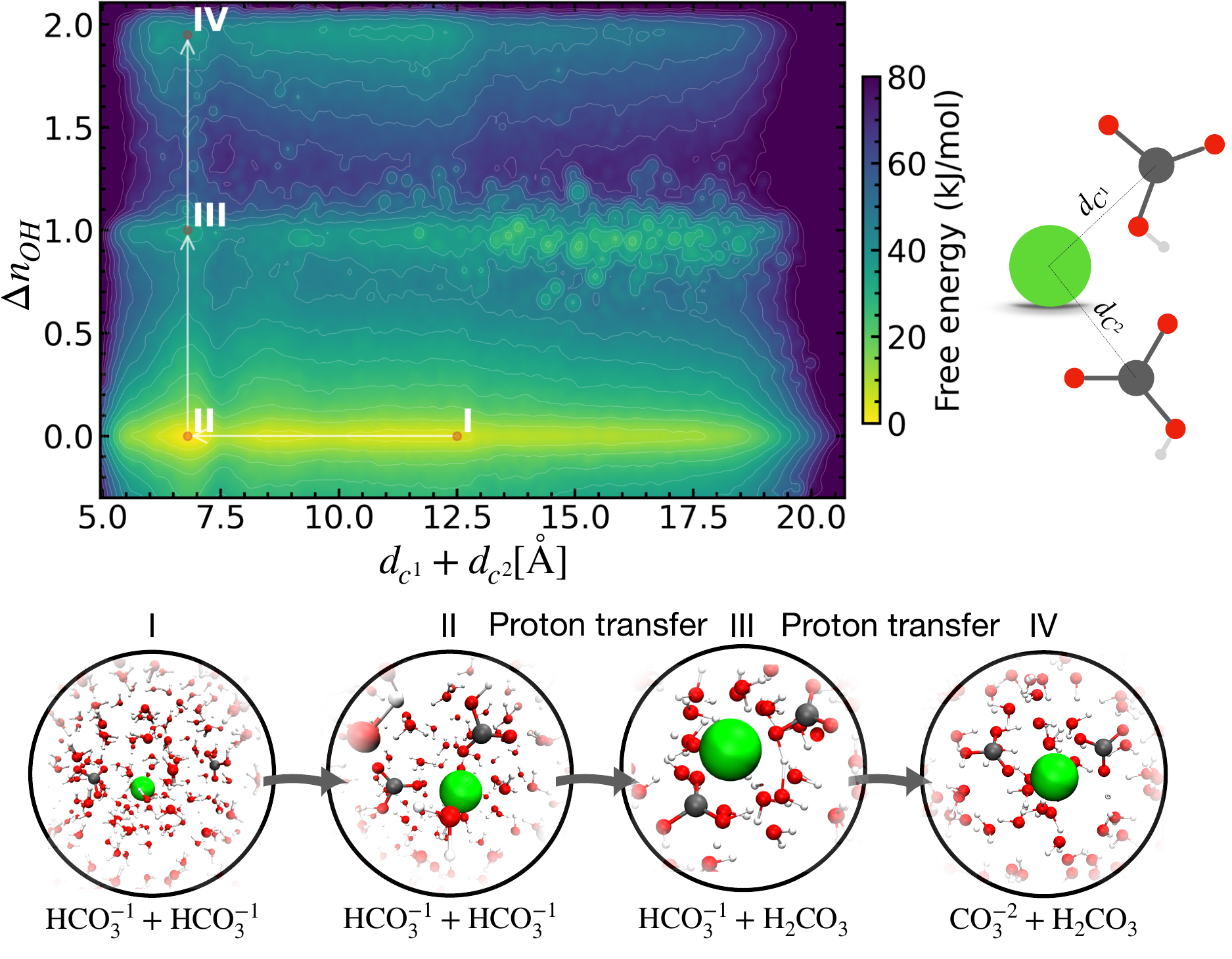}
\caption{
Free energy landscape of Ca$^{2+}$ association with two bicarbonate ions in aqueous solution.
The free energy $F(d_{C^{1}}+d_{C^{2}},\Delta n_{\mathrm{OH}})$ (color scale) is shown as a function of the sum of the two Ca--C distances, $d_{C^{1}}+d_{C^{2}}$, and the protonation coordinate $\Delta n_{\mathrm{OH}} = |n_{\mathrm{OH}}^{(1)} - n_{\mathrm{OH}}^{(2)}|$, which measures the difference in protonation between the two bicarbonate ions.
Relevant states are marked with Roman numerals and representative configurations are shown below.
}
    \label{fig:ca-two-bicarb}
\end{figure}

We next examine a three-ion case: a single Ca$^{2+}$ ion interacting with two bicarbonate ions.
Figure~\ref{fig:ca-two-bicarb} shows the free-energy surface projected onto a structural coordinate—the sum of the two Ca--C distances, $d_{C^{1}}+d_{C^{2}}$—and a chemical coordinate $\Delta n_{\mathrm{OH}}$, which measures the difference in protonation between the two bicarbonate ions (see Methods section for details).
The landscape reveals distinct basins corresponding to configurations with two bicarbonate ions (I–II), one bicarbonate and one carbonic acid (III), and one carbonate and one carbonic acid (IV).
In Figure~\ref{fig:ca-two-bicarb} we highlight the pathway for the transition from the starting configuration (I) with two bicarbonate ions separated from Ca$^{2+}$ by a relatively large distance to a final state (IV) with one carbonate and one carbonic acid in contact with Ca$^{2+}$.
Along this pathway, first the two bicarbonate ions come in close contact with Ca$^{2+}$, corresponding to configuration II in Figure~\ref{fig:ca-two-bicarb}.
Note that the close proximity between Ca$^{2+}$ and bicarbonates is essential for the process to occur.
Then, in the configuration labeled III a proton transfer event from a neighboring water molecule to one of the bicarbonates leads to the formation of one carbonic acid molecule.
Finally, from III to IV the remaining bicarbonate ion transforms into carbonate via a Grotthuss-type mechanism, i.e., through a sequence of proton jumps through the hydrogen bond network ~\cite{marx2006proton,hassanali2013proton,agmon2016protons}.
As in the case of the ion pair, we find here that ion association facilitates proton transfer processes and is a key enabler of species interconversion.
Note that state IV has a high free energy and we hypothesize that it evolves to a more stable configuration through the conversion of carbonic acid into \ce{CO2}.
Block and cumulative analyses showed that the principal free-energy basins and proton-transfer regions were preserved throughout the production trajectories for both systems. Higher-free-energy regions displayed greater statistical uncertainty. The free-energy surfaces are therefore considered sufficiently converged to support the proposed mechanistic interpretation, while no uniformly precise quantitative barrier heights are inferred. The corresponding block and cumulative convergence analyses, including the analysis notebooks and associated data, are provided in the repository accompanying this work.

\begin{figure*}[hbt!]
    \centering
    \includegraphics[width=\textwidth]{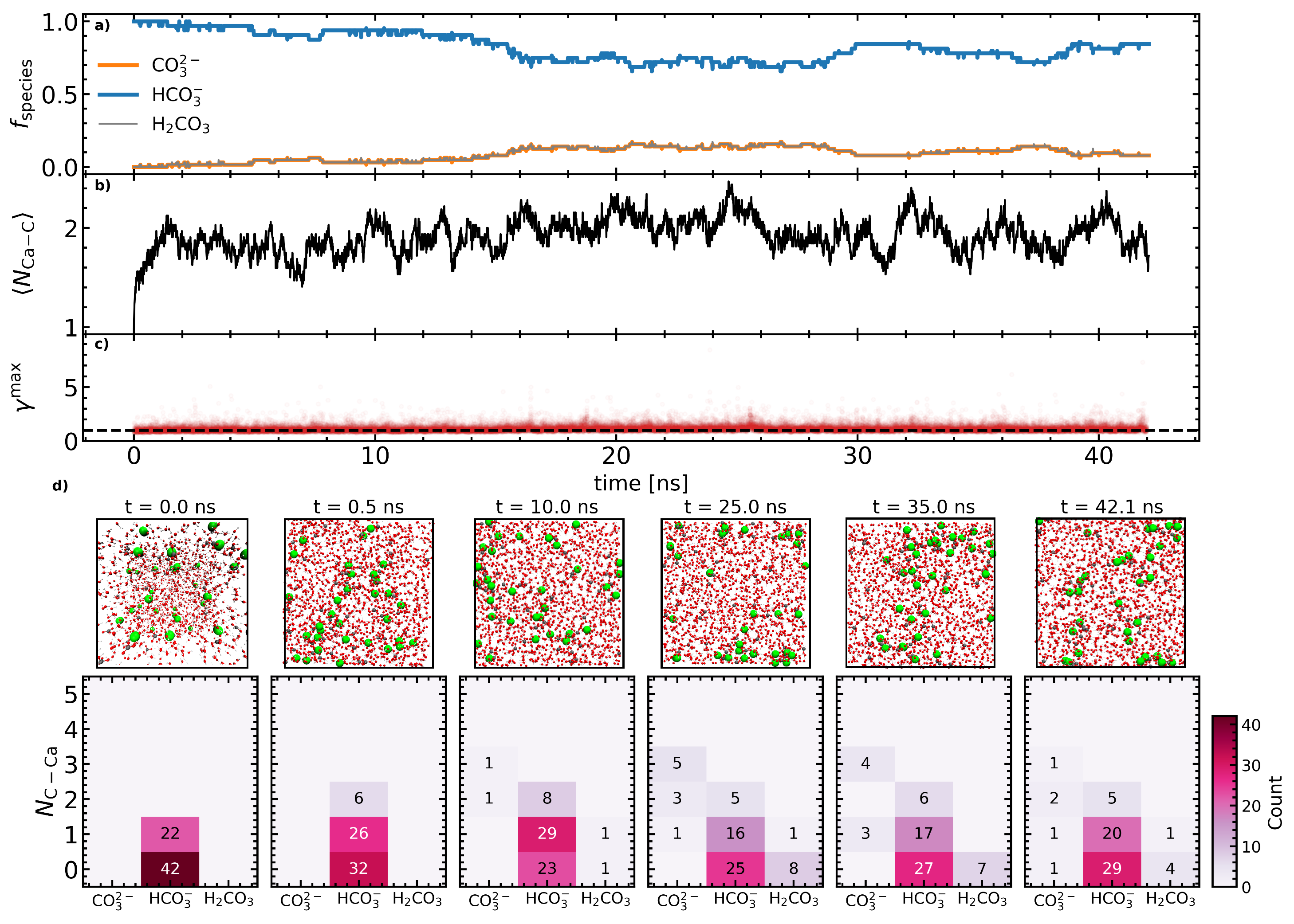}
    \caption{
Initial stages of \ce{CaCO3} precipitation at near-neutral pH from unbiased molecular dynamics.
(a) Fraction of \ce{CO3^2-}, \ce{HCO3-}, and \ce{H2CO3} species, showing interconversion on nanosecond timescales.
(b) Average coordination number of carbonate species around Ca, $\langle N_{Ca-C} \rangle$, indicating ion association.
(c) Maximum ACE extrapolation grade $\gamma^\mathrm{max}$ monitored throughout the trajectory. Values close to one indicate that the sampled local environments remain within the interpolation domain represented by the training data.
(d) For each carbonate, we show the coordination number with Ca and the ion state (\ce{CO3^2-}, \ce{HCO3-}, or \ce{H2CO3}) at six 
representative times. We also show the corresponding atomistic configurations. The coordination numbers used for the analysis in panels (b) and (d) are discrete counts computed using a 4~\AA\ cutoff; they are distinct from the smooth coordination-number CV defined in Eqs.~(1) and (2).
}
    \label{fig:species}
\end{figure*}

Having shown from biased simulations that ion association enables proton transfer in two- and three-ion systems, we now examine how these processes occur in many-ion solutions.
To this end, we performed long unbiased molecular dynamics simulations at near-neutral pH, starting from solutions of Ca$^{2+}$ and bicarbonate ions, and using approximately 5000 atoms.
Figure~\ref{fig:species}a shows the time evolution of the \ce{CO3^2-}, \ce{HCO3-}, and \ce{H2CO3} fractions.
The figure reveals continuous interconversion between protonation states over nanosecond timescales, with all species remaining populated throughout the trajectory.
Figure~\ref{fig:species}b shows the average  coordination number of carbonate species around Ca, $\langle N_{Ca-C} \rangle$, as a measure of ion association.
The coordination number increases during the first few nanoseconds, indicating the formation of associated ion configurations, and then fluctuates around a steady value $\langle N_{Ca-C} \rangle \approx 2$.
Considering the complexity of the environments that appear during this simulation, it is important to assess the ability of our ACE potential to describe them.
For this purpose, we monitored the maximum extrapolation grade $\gamma^\mathrm{max}$, shown in Figure~\ref{fig:species}c. Its values remained close to one, indicating that the sampled local environments remained within the interpolation domain represented by the training data.

We then analyzed the interplay between chemical reactivity and ion association for individual carbonate ions.
In Figure~\ref{fig:species}d, for each carbonate, we show the coordination number with Ca and the ion state (\ce{CO3^2-}, \ce{HCO3-}, or \ce{H2CO3}) at different times during the simulation.
We observe that initially \ce{HCO3-} ions are coordinated by none or one Ca ion, but during the initial transient ($t=0.5$ ns) some of them become coordinated by 2\,Ca ions.
Afterwards, at around $t=10$ ns, some of the \ce{HCO3-} have transformed into \ce{CO3^2-}, and \ce{H2CO3}.
Moreover, we observe that \ce{CO3^2-} ions are coordinated by 2 to 3\,Ca ions, while \ce{H2CO3} is mostly non associated to Ca.
Thus, \ce{H2CO3} leaves the Ca environment after the \ce{HCO3-} to \ce{H2CO3} reaction takes place.
It is likely that \ce{H2CO3} would subsequently decompose into CO$_2$ and water, yet we do not observe this phenomenon within the relatively short time of our simulation.
Figure~\ref{fig:species}d clearly shows the trend \ce{CO3^2-} $\textgreater$ \ce{HCO3-} $\textgreater$ \ce{H2CO3} in the coordination of these species with Ca.
The atomistic configurations depicted in Figure~\ref{fig:species}d also show a limited tendency to ion aggregation, consistent with Figure~\ref{fig:species}b.

\begin{figure*}[hbt!]
    \centering
    \includegraphics[width=\textwidth]{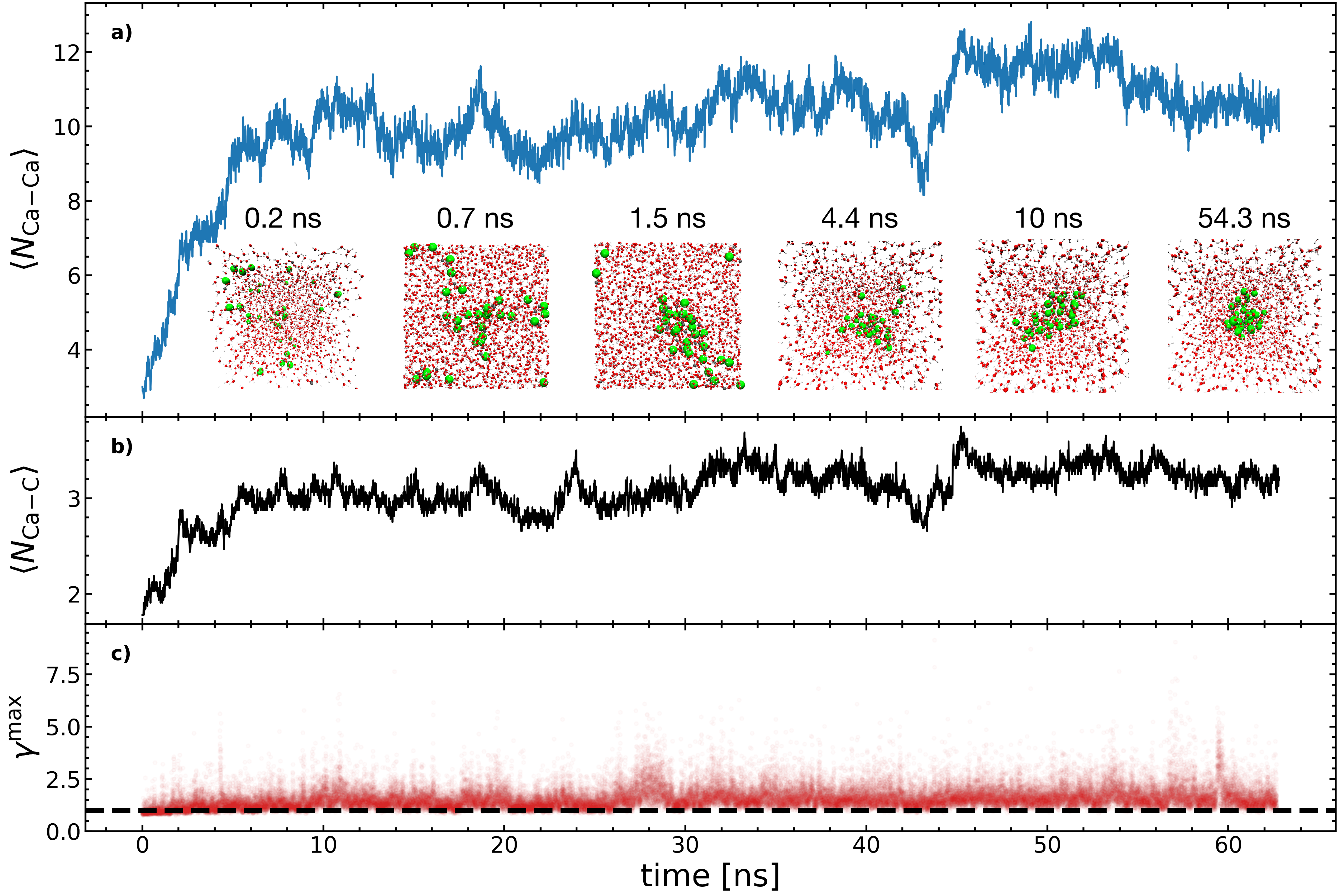}
    \caption{
Initial stages of \ce{CaCO3} precipitation at high-pH conditions from unbiased molecular dynamics.
(a) Time evolution of the average Ca--Ca coordination number showing rapid cluster formation. Insets depict representative configurations at increasing times.
(b) Ca--C coordination during aggregation.
(c) Maximum ACE extrapolation grade, $\gamma^\mathrm{max}$, monitored throughout the trajectory. Values close to one indicate that the sampled local environments remain within the interpolation domain represented by the training data.
}
    \label{fig:aggregation}
\end{figure*}

Having established how ion association and speciation are coupled in solution, we now examine how these dynamics evolve during aggregation under high-pH conditions, where carbonate species remain predominantly in the \ce{CO3^2-} form.
Figure~\ref{fig:aggregation}a shows the time evolution of the Ca--Ca coordination number, which characterizes the formation of  clusters.
The coordination number increases rapidly within the first $\sim$8\,ns and reaches a plateau after approximately 10\,ns, indicating the formation of an aggregated structure that persists over the remainder of the trajectory.
Representative configurations shown in the figure confirm the transition from dispersed ions to a compact cluster.
Figure~\ref{fig:aggregation}b shows the Ca--C coordination during this process.
The coordination remains finite throughout the trajectory, indicating that carbonate ions remain associated with Ca$^{2+}$ ions within the aggregated structure.
In contrast to the near-neutral pH case, no changes in chemical speciation are observed during aggregation.
Carbonate species remain in the \ce{CO3^2-} form throughout the simulation, and no proton-transfer events are observed over the simulated timescale. Because protons are available only through water molecules in this setup, the absence of proton transfer may partly reflect the high free-energy cost of water deprotonation.
Overall, the figure shows that aggregation proceeds without an observed change in carbonate speciation under the simulated carbonate-rich conditions, consistent with experimental observations~\citenum{jin2025formation}.

The current simulation framework has important advantages and limitations worth considering.
The use of revPBE-D3, a computationally efficient density functional with dispersion corrections, allowed us to construct a training dataset of sufficient diversity to describe many-ion systems, reactive intermediates, and interfacial environments within a single model.
This level of chemical and configurational diversity would not be currently achievable with more computationally demanding electronic structure methods.
Recent work has employed coupled-cluster CCSD(T) and other correlated wavefunction methods to study single ion-pair association with greater quantitative accuracy\cite{oneill2026ccsd,bian2026transfer}.
While such approaches provide a valuable benchmark for thermodynamic properties, it remains unclear how to extend them to the chemically complex, many-ion environments studied here, which are directly relevant to nucleation and mineral carbonation.
In this sense, the present methodology occupies a complementary position: it sacrifices some quantitative accuracy at the few-ion level in order to access the collective regime.
Future work could address this trade-off by incorporating more accurate electronic structure references for key reactive pathways, or by using the present model to identify the configurations where high-level benchmarks are most needed.
Additional future research directions include the explicit treatment of nuclear quantum effects, which may influence proton-transfer rates and hydrogen-bond dynamics, and the incorporation of long-range electrostatic interactions beyond the local cutoff of the ACE framework, which could improve quantitative predictions for ion-pairing thermodynamics and cluster stability in dilute solution.
We note that Ref.~\citenum{piaggi2025ab} shows that long-range interactions affect ion association free energy curves only quantitatively and that short range models are able to properly capture the ion association mechanism.
All simulation cells were overall charge neutral. Proton transfer and the associated changes in local bonding are represented implicitly through the learned potential-energy surface; however, the model contains no explicit electronic charges or long-range charge-transfer degrees of freedom. In addition, the approximately 1~M concentrations used here are substantially higher than those typical of seawater and were chosen to make aggregation observable within accessible simulation times. The many-ion results should therefore be interpreted as mechanistic observations for concentrated solutions rather than as quantitative predictions for seawater conditions.

In conclusion, we have developed a fully reactive machine-learning potential for aqueous calcium carbonate based on the ACE framework and the revPBE-D3 functional, enabling molecular dynamics simulations of many-ion systems with near first-principles accuracy. 
By combining this potential with enhanced sampling and large-scale unbiased simulations, we have established a mechanistic picture of early-stage calcium carbonate formation that shows a rich interplay between molecular reactivity and ion aggregation.
A central finding of this work is that ion association provides favorable pathways for proton transfer, enabling chemical reactivity that was not observed for isolated ions in solution.
In unbiased simulations of many-ion systems, we observe distinct behavior at near-neutral and high pH.
At near-neutral pH, where bicarbonate ions dominate, simulations show limited tendency to ion aggregation, continuous interconversion between species, and significant changes in the coordination environments as species evolve.
At high pH, where carbonate species predominate, aggregation is rapid and is not reversed over the simulated timescale, while carbonate speciation remains unchanged and no proton-transfer events are observed.
These results demonstrate that proton transfer (chemical speciation) and structural aggregation are not independent processes: the local chemical environments generated during aggregation determine barriers for proton transfer, and the protonation state of ions in turn determine how they associate.
More broadly, this work shows that machine-learning-driven ab initio simulations together with enhanced sampling can be used to better understand reactive crystallization processes in complex aqueous environments, which are important to biomineralization, carbon sequestration through mineral carbonation, and the formation of cement-based materials.

\section*{Computational methods}
\label{sec:methods}

\subsection*{Reference electronic-structure calculations}

We used revPBE-D3 to recompute and extend the dataset for calcium carbonate in aqueous solution reported in Ref.~\citenum{piaggi2025ab}.
Our new dataset contains diverse CaCO$_3$ environments, including ion pairing, proton-transfer states, multi-ion clusters, multiple carbonate species, and crystalline calcium carbonate polymorphs.
Reference energies and forces for training and active-learning refinement were obtained from plane-wave DFT calculations performed with the \textsc{Quantum ESPRESSO} package v6.4.1 ~\cite{giannozzi2009quantum,giannozzi2017advanced}.
The exchange–correlation energy was described using the revised Perdew–Burke–Ernzerhof (revPBE) generalized gradient approximation~\cite{perdew1996generalized,zhang1998comment} with D3 dispersion correction and Becke–Johnson damping~\cite{grimme2010consistent,grimme2011effect}.
This functional was selected to ensure full consistency with our previously developed general-purpose ACE potential for water~\cite{ibrahim2024efficient,ibrahim2026pd}, which demonstrated accurate reproduction of liquid structure, ice polymorph energetics, and the water phase diagram across a broad thermodynamic range.
We employed norm-conserving, scalar-relativistic
pseudopotentials~\cite{hamann2013optimized} for Ca, C, O, H parameterized using the PBE~\cite{perdew1996generalized}
functional with 10\,, 4\,, 6\,, and 1\, valence electrons, respectively.
A kinetic-energy cutoff of 110\,Ry was used for the plane-wave expansion of the wavefunctions and 440\,Ry for the charge density.
For the disordered aqueous configurations considered here, $\Gamma$-point sampling was sufficient to converge energies and forces within the target accuracy of the machine-learning model.
Electronic self-consistency was achieved using a convergence threshold of $10^{-8}$\,Ry in total energy.

\subsection*{ACE Potential and Training Procedure}

To train a transferable model for this chemically diverse system, we employed the atomic cluster expansion (ACE) formalism~\cite{drautz2019atomic}.
ACE provides a formally complete, systematically convergent, and physically interpretable representation of local atomic environments.
We employed a shifted-and-scaled Finnis--Sinclair embedding with a nonlinear representation of the atomic energy based on two atomic properties, each represented by an ACE basis expansion~\cite{drautz2019atomic,lysogorskiy2021performant,bochkarev2022efficient}.

The training dataset comprised 25,819 configurations containing 17,208,893\,atomic environments, including solvated ions, proton-transfer states, multicarbonate clusters, and crystalline polymorphs.
The model parameters were optimized using the BFGS algorithm with uniform configuration weighting and a relative force weight of $\kappa=0.33$ in the combined energy--force loss function.
Weak $L_1$ and $L_2$ regularization of the expansion coefficients and radial smoothness regularization were applied.

The ACE basis was truncated at fifth body order with a cutoff radius of 6.0\,\AA. A Chebyshev polynomial radial basis was employed with $n=[15,3,2,1,1]$ and $l=[0,2,2,1,1]$, where successive entries correspond to increasing body order.
The resulting potential contains 2,064\,basis functions per element, corresponding to 8,256\,basis functions and 17,862\,parameters in total.
All ACE parameterizations were performed using the PACEmaker package~\cite{lysogorskiy2021performant,bochkarev2022efficient}.

Active learning was performed iteratively using the D-optimality criterion~\cite{lysogorskiy2023active,podryabinkin2017active}. Configurations with an extrapolation grade $\gamma>10$ were identified as extrapolative with respect to the current training domain, and their reference energies and forces were calculated at the DFT level before being added to the training dataset.

\subsection*{Enhanced-sampling simulations}
\label{sec:opes}

To generate diverse configurations representative of reactive aggregation of \ce{CaCO3} and to compute free-energy landscapes of ion association, we performed enhanced-sampling simulations using the OPES method~\cite{invernizzi2020rethinking,invernizzi2020unified} in its variant that targets the well-tempered ensemble.
OPES was used to construct a bias potential as a function of the collective variables (CVs) described below.
Statistical uncertainties in the OPES-reweighted free-energy surfaces were assessed by dividing each equilibrated production trajectory into six equal contiguous blocks and reconstructing the free-energy surface independently for each block. Confidence intervals at the 95\% level were estimated from the block-to-block variability using the Student-\(t\) distribution. Convergence was additionally assessed using cumulative analyses based on 25\%, 50\%, 75\%, and 100\% of each production trajectory.

\paragraph{Active learning simulations.}

During active learning, the OPES simulations used two collective variables aimed at promoting both ion aggregation and reactivity. The first is the mean Ca-C coordination number defined through the formula,
\begin{equation}
    \langle N_\mathrm{Ca-C} \rangle=\frac{1}{N_\mathrm{Ca}}\sum\limits_{i \in \mathrm{Ca}}\sum\limits_{j \in \mathrm{C}} s(r_{ij}),    
\end{equation}
where $r_{ij}$ is the distance between Ca atom $i$ and C atom $j$, $N_\mathrm{Ca}$ is the total number of Ca atoms, and $s(r)$ is a smooth cubic switching function,
\begin{equation}
s(r)=
\begin{cases}
1, & r\le r_0,\\
1-3x(r)^2+2x(r)^3, & r_0<r<r_{\max},\\
0, & r\ge r_{\max},
\end{cases}
\end{equation}
with $x(r)=(r-r_0)/(r_{\max}-r_0)$, using $r_0 = 0.33$~nm and $r_{\max} = 0.41$~nm.
The values for $r_0$ and $r_{\max}$ were chosen based on the radial distribution functions and aim at capturing bond formation/breaking.
This smooth coordination number was used as an OPES collective variable during active learning. The coordination numbers used to analyze the unbiased trajectory in Fig.~\ref{fig:species} were instead evaluated as discrete counts with a 4~\AA\ cutoff.

The second biased collective variable was the number of carbonate species in the simulation box. To construct this variable, we first defined the coordination number between O atoms in a given carbonate $k$ and all H atoms,
\begin{equation}
    n_{\mathrm{OH}}^k=\sum\limits_{i \in \mathrm{O}_\mathrm{C}}\sum\limits_{j \in \mathrm{H}} s(r_{ij}),
    \label{eq:n_OH}
\end{equation}
where O$_\mathrm{C}$ are O atoms in carbonate $k$, $r_{ij}$ is the distance between O$_\mathrm{C}$ atom $i$ and H atom $j$, and $s(r)$ is a smooth cubic switching function with $r_0 = 0.095$~nm and $r_{\max} = 0.16$~nm.
$n_{\mathrm{OH}}^k$ can have values of 0, 1, or 2, if the molecule is \ce{CO3^2-}, \ce{HCO3-}, or \ce{H2CO3}, respectively.
We also define the coordination number of the C atom in carbonate $k$ and O atoms,
\begin{equation}
    n_{\mathrm{C}\mathrm{O}}^k=\sum\limits_{j \in \mathrm{O}} s(r_{kj}),
\end{equation}
where $r_{kj}$ is the distance between C atom in carbonate $k$ and O atom $j$, and $s(r)$ is a smooth cubic switching function with $r_0 = 0.13$~nm and $r_{\max} = 0.34$~nm.
$n_{\mathrm{C}\mathrm{O}}^k$ can have values of 2 or 3, if the molecule is \ce{CO2} or carbonate-like, respectively.
We can now define the number of carbonate (\ce{CO3^2-}) species using,
\begin{equation}
    n_{\mathrm{CO}_3^{2-}}=\sum\limits_{k \in \mathrm{C}} e^{-\frac{\left(n_{\mathrm{OH}}^k\right)^2}{2 \: \sigma^2}} e^{-\frac{\left(n_{\mathrm{C}\mathrm{O}}^k-3\right)^2}{2 \: \sigma^2}},
\end{equation}
where $k \in \mathrm{C}$ is a sum over all C atoms and we used $\sigma=0.3$.
We also monitored other species, such as \ce{HCO3-}, \ce{H2CO3} and \ce{CO2} using similar formulae.

The OPES simulations used $\langle N_\mathrm{Ca-C} \rangle$ and $n_{\mathrm{CO}_3^{2-}}$ as collective variables, a maximum bias barrier of 80~kJ~mol$^{-1}$, and were aimed at generating diverse configurations rather than at converging the free energy surface.

\paragraph{Ca$^{2+}$–carbonate system.}

For the Ca$^{2+}$–CO$_3^{2-}$ ion pair in aqueous solution, the free-energy surface was constructed as a function of two collective variables: (i) the Ca–C distance, describing ion association, and (ii) the coordination number between carbonate oxygen atoms and protons, $n_\mathrm{OH}$, capturing protonation states. The coordination number is defined as in Eq.~\eqref{eq:n_OH} above. OPES simulations were run at temperature $T=330$~K, with the bias updated every 500 time steps and a maximum bias barrier of 50~kJ~mol$^{-1}$.
Upper-wall restraints were applied to limit unphysical ion separation and finite-size effects.

\paragraph{Ca$^{2+}$–two-bicarbonate system.}

For the three-ion system consisting of Ca$^{2+}$ and two HCO$_3^{-}$ ions, two collective variables were employed: (i) the sum of the two Ca–C distances, $d_1 + d_2$, describing overall ion association, and (ii) the absolute difference in O–H coordination numbers ($n_{\mathrm{OH}}^k$) between the two bicarbonate groups, capturing proton transfer and asymmetry in protonation.
Additional harmonic restraints were applied to preserve physically meaningful carbonate geometries, ensuring that each carbon remains coordinated to at most three oxygen atoms.
Simulations were performed at $T=330$~K with OPES bias updated every 500 steps and a maximum barrier of 60~kJ~mol$^{-1}$. Upper-wall restraints were used to control ion separation and coordination.
The complete definitions and parameters of the upper walls and harmonic restraints are given in the PLUMED input files that will be provided with the simulation and analysis repository.

\paragraph{Free-energy reconstruction.}

Unbiased free-energy surfaces were obtained by reweighting the biased trajectories. The free energy is defined as
\begin{equation}
F(\mathbf{s}) = -k_{\mathrm{B}}T \ln P(\mathbf{s}),
\end{equation}
where $\mathbf{s}$ denotes the set of collective variables.

The The unbiased probability distribution was computed as
\begin{equation}
P(\mathbf{s})
=
\frac{
\left\langle
\delta\!\left(\mathbf{s}-\mathbf{s}(\Gamma)\right)
e^{\beta V(\mathbf{s})}
\right\rangle_V
}{
\left\langle e^{\beta V}\right\rangle_V
},
\end{equation}
where $\langle\cdot\rangle_V$ denotes an ensemble average in the presence of the bias potential $V$.

\subsection*{Molecular dynamics}

All simulations were performed using LAMMPS~\cite{LAMMPS} with the PACE~\cite{lysogorskiy2021performant} implementation of the ACE potential.
Short-range core repulsion was modeled using the Ziegler–Biersack–Littmark (ZBL) screened nuclear potential for interatomic distances below the range represented in the training dataset~\cite{ziegler1985srim}.
Periodic boundary conditions were applied in all directions, and a time step of $\Delta t = 0.5$~fs was used.

Biased simulations were carried out using PLUMED~2.9.2~\cite{tribello2014plumed,plumed2019promoting} interfaced with LAMMPS. For the two-ion system, simulations were performed for 15\,ns in a box containing 197 water molecules, with a maximum OPES bias of 50~kJ~mol$^{-1}$.
For the three-ion system, simulations were extended to 60\,ns in a box of 192 water molecules with a 60~kJ~mol$^{-1}$ bias.

Unbiased simulations were initialized from bicarbonate-only and carbonate-only solutions, hereafter termed ``near-neutral'' and ``high-pH'' conditions, respectively, according to their initial carbonate speciation. These labels reflect the carbonate acid--base equilibrium, \ce{HCO3- <=> CO3^2- + H+}, for which the pH is related to carbonate speciation through the Henderson--Hasselbalch equation, $\mathrm{pH}=pK_{a,2}+\log_{10}\!\left(a_{\mathrm{CO_3^{2-}}}/a_{\mathrm{HCO_3^-}}\right)$, where $a_i$ denotes the activity of species $i$. Thus, a larger carbonate-to-bicarbonate activity ratio corresponds to higher pH. The pH was not imposed or calculated explicitly; the terms ``near-neutral'' and ``high-pH'' refer to the initial bicarbonate-rich and carbonate-rich compositions, respectively.
We employed bicarbonate and carbonate concentrations of approximately 1.01~M (4768 atoms) and 1.03~M (4720 atoms), respectively.
The bicarbonate system contained 29~\ce{Ca^2+}, 58~\ce{HCO3-}, and 1483~\ce{H2O}, whereas the carbonate system contained 29~\ce{Ca^2+}, 29~\ce{CO3^2-}, and 1525~\ce{H2O}; both systems were charge neutral.

Temperature was maintained at $T=330$~K using a canonical velocity-rescaling thermostat ($\tau_T=0.1$~ps), and pressure was controlled at $P=1.013$~bar using an isotropic barostat ($\tau_P=1.0$~ps).

\section*{Data availability}
All data and files supporting this work, including the DFT reference dataset, ACE potential and training configuration, molecular dynamics and enhanced sampling input files, and analysis notebooks, are openly available on Zenodo~\cite{ibrahim2026reactive_dataset,ibrahim2026reactive_repo}.

\section*{Acknowledgements}
E.I acknowledges funding through the International Max Planck Research School for Sustainable Metallurgy (IMPRS SusMet).
We acknowledge computational resources from the Red Española de Supercomputación resources provided by Barcelona Supercomputing Center in MareNostrum to RES-FI-2024-2-0026.
We gratefully acknowledge the computing time provided on the high-performance computing system Noctua~2 at the NHR Center Paderborn for Parallel Computing (PC$^2$), under project ID~4806.
The NHR Center PC$^2$ is jointly supported by the Federal Ministry of Education and Research and the state governments participating in the National High-Performance Computing (NHR) joint funding program.

\bibliography{refs.bib}

@misc{ibrahim2026reactive_repo,
  author    = {Ibrahim, Eslam and Lysogorskiy, Yury and Drautz, Ralf and Piaggi, Pablo M.},
  title     = {Reactive calcium carbonate precipitation with {ACE}: Analysis code and simulation inputs},
  publisher = {Zenodo},
  year      = {2026},
  doi       = {10.5281/zenodo.22238939},
  url       = {https://doi.org/10.5281/zenodo.22238939}
}

@misc{ibrahim2026reactive_dataset,
  author    = {Ibrahim, Eslam and Lysogorskiy, Yury and Drautz, Ralf and Piaggi, Pablo M.},
  title     = {Reactive calcium carbonate precipitation from an atomic cluster expansion potential and enhanced sampling: Dataset},
  publisher = {Zenodo},
  year      = {2026},
  doi       = {10.5281/zenodo.22238290},
  url       = {https://doi.org/10.5281/zenodo.22238290}
}

@article{zhang1998comment,
  author  = {Zhang, Yingkai and Yang, Weitao},
  title   = {Comment on {``Generalized Gradient Approximation Made Simple''}},
  journal = {Physical Review Letters},
  year    = {1998},
  volume  = {80},
  number  = {4},
  pages   = {890--890},
  doi     = {10.1103/PhysRevLett.80.890}
}

@book{ziegler1985srim,
  author    = {Ziegler, James F. and Biersack, Jochen P. and Littmark, Uffe},
  title     = {The Stopping and Range of Ions in Solids},
  publisher = {Pergamon Press},
  address   = {New York},
  year      = {1985}
}

@article{kohn-1965,
  author  = {Kohn, W. and Sham, L. J.},
  journal = {Phys. Rev. },
  month   = {11},
  number  = {4A},
  pages   = {A1133--A1138},
  title   = {{Self-Consistent equations including exchange and correlation effects}},
  volume  = {140},
  year    = {1965},
}

@article{jin2025formation,
  title={Formation, chemical evolution and solidification of the dense liquid phase of calcium (bi) carbonate},
  author={Jin, Biao and Chen, Ying and Pyles, Harley and Baer, Marcel D and Legg, Benjamin A and Wang, Zheming and Washton, Nancy M and Mueller, Karl T and Baker, David and Schenter, Gregory K and others},
  journal={Nat. Mater.},
  volume={24},
  number={1},
  pages={125--132},
  year={2025},
  publisher={Nature Publishing Group UK London}
}

@article{hassanali2013proton,
  title={Proton transfer through the water gossamer},
  author={Hassanali, Ali and Giberti, Federico and Cuny, J{\'e}r{\^o}me and K{\"u}hne, Thomas D and Parrinello, Michele},
  journal={Proc. Natl. Acad. Sci.},
  volume={110},
  number={34},
  pages={13723--13728},
  year={2013},
  publisher={National Academy of Sciences}
}

@article{de2015crystallization,
  title={Crystallization by particle attachment in synthetic, biogenic, and geologic environments},
  author={De Yoreo, James J and Gilbert, Pupa UPA and Sommerdijk, Nico AJM and Penn, R Lee and Whitelam, Stephen and Joester, Derk and Zhang, Hengzhong and Rimer, Jeffrey D and Navrotsky, Alexandra and Banfield, Jillian F and others},
  journal={Science},
  volume={349},
  number={6247},
  pages={aaa6760},
  year={2015},
  publisher={American Association for the Advancement of Science}
}

@article{di2009theoretical,
  title={Theoretical study of the dimerization of calcium carbonate in aqueous solution under natural water conditions},
  author={Di Tommaso, Devis and de Leeuw, Nora H},
  journal={Geochim. Cosmochim. Acta.},
  volume={73},
  number={18},
  pages={5394--5405},
  year={2009},
  publisher={Elsevier}
}

@article{smeets2017classical,
  title={A classical view on nonclassical nucleation},
  author={Smeets, Paul JM and Finney, Aaron R and Habraken, Wouter JEM and Nudelman, Fabio and Friedrich, Heiner and Laven, Jozua and De Yoreo, James J and Rodger, P Mark and Sommerdijk, Nico AJM},
  journal={Proc. Natl. Acad. Sci.},
  volume={114},
  number={38},
  pages={E7882--E7890},
  year={2017},
  publisher={National Academy of Sciences}
}

@article{raiteri2010water,
  title={Water is the key to nonclassical nucleation of amorphous calcium carbonate},
  author={Raiteri, Paolo and Gale, Julian D},
  journal={J. Am. Chem. Soc.},
  volume={132},
  number={49},
  pages={17623--17634},
  year={2010},
  publisher={ACS Publications}
}

@article{zhu2025molecular,
  title={Molecular mechanisms of CO2 mineralization on wetting nanoscale surfaces using molecular simulations and metadynamics},
  author={Zhu, Xinping and Tao, Yong and Dupuis, Romain and Gao, Yining and Poon, Chi-Sun and Ioannidou, Katerina and Pellenq, Roland JM},
  journal={Nat. Commun.},
  volume={16},
  number={1},
  pages={10758},
  year={2025},
  publisher={Nature Publishing Group UK London}
}

@article{bian2026transfer,
      author = {Bian, Xuezhi and Carter, Emily A.},
    title = {Transfer Learning
Meets Embedded Correlated Wavefunction
Theory for Chemically Accurate Molecular Simulations: Application
to Calcium Carbonate Ion Pairing},
    journal = {J. Chem. Theory Comput.},
    volume = {22},
    number = {10},
    pages = {5174-5184},
    year = {2026},
    month = {05},
     issn = {1549-9618},
    doi = {10.1021/acs.jctc.6c00403},
    url = {https://doi.org/10.1021/acs.jctc.6c00403},
    eprint = {https://pubs.acs.org/jctcce/article-pdf/22/10/5174/66178826/acs.jctc.6c00403.pdf}
}

@misc{oneill2026ccsd,
      title={From Accurate Quantum Chemistry to Converged Thermodynamics for Ion Pairing in Solution}, 
      author={Niamh O'Neill and Benjamin X. Shi and William C. Witt and Blake I. Armstrong and William J. Baldwin and Paolo Raiteri and Christoph Schran and Angelos Michaelides and Julian D. Gale},
      year={2026},
      eprint={2603.06800},
      archivePrefix={arXiv},
      primaryClass={physics.chem-ph},
}

@article{li2024ion,
  title={Ion association behaviors in the initial stage of calcium carbonate formation: An ab initio study},
  author={Li, Yue and Zhang, Jiarui and Zeng, Hongbo and Zhang, Hao},
  journal={J. Chem. Phys.},
  volume={161},
  number={1},
  year={2024},
  publisher={AIP Publishing}
}

@article{tommaso2008onset,
  title={The onset of calcium carbonate nucleation: a density functional theory molecular dynamics and hybrid microsolvation/continuum study},
  author={Tommaso, Devis Di and De Leeuw, Nora H},
  journal={J. Phys. Chem. B},
  volume={112},
  number={23},
  pages={6965--6975},
  year={2008},
  publisher={ACS Publications}
}

@article{wu2024bicarbonate,
  title={Bicarbonate-mediated proton transfer requires cations},
  author={Wu, Qianbao and Yang, Na and Xiao, Mengjun and Wang, Wei and Cui, Chunhua},
  journal={Nat. Commun.},
  volume={15},
  number={1},
  pages={9145},
  year={2024},
  publisher={Nature Publishing Group UK London}
}

@article{ibrahim2026pd,
        title={Water Phase Diagram from a General-Purpose Atomic Cluster Expansion Potential},
  author={Ibrahim, Eslam and Lysogorskiy, Yury and Drautz, Ralf and Piaggi, Pablo M},
journal = {J. Chem. Theory Comput.},
  volume={22},
  number={9},
  pages={4758--4766},
  year={2026},
  publisher={ACS Publications},
  doi={10.1021/acs.jctc.6c00094}
}

@article{henzler2018supersaturated,
  title={Supersaturated calcium carbonate solutions are classical},
  author={Henzler, Katja and Fetisov, Evgenii O and Galib, Mirza and Baer, Marcel D and Legg, Benjamin A and Borca, Camelia and Xto, Jacinta M and Pin, Sonia and Fulton, John L and Schenter, Gregory K and others},
  journal={Science advances},
  volume={4},
  number={1},
  pages={eaao6283},
  year={2018},
  publisher={American Association for the Advancement of Science}
}

@article{tribello2009molecular,
  title={A molecular dynamics study of the early stages of calcium carbonate growth},
  author={Tribello, Gareth A and Bruneval, Fabien and Liew, CheeChin and Parrinello, Michele},
  journal={J. Phys. Chem. B},
  volume={113},
  number={34},
  pages={11680--11687},
  year={2009},
  publisher={ACS Publications}
}

@article{huang2021uncovering,
  title={Uncovering the role of bicarbonate in calcium carbonate formation at near-neutral pH},
  author={Huang, Yu-Chieh and Rao, Ashit and Huang, Shing-Jong and Chang, Chun-Yu and Drechsler, Markus and Knaus, Jennifer and Chan, Jerry Chun Chung and Raiteri, Paolo and Gale, Julian D and Gebauer, Denis},
  journal={Angew. Chem. Int. Ed.},
  volume={60},
  number={30},
  pages={16707--16713},
  year={2021},
  publisher={Wiley Online Library}
}

@article{hamann2013optimized,
  title={Optimized norm-conserving Vanderbilt pseudopotentials},
  author={Hamann, Don R},
  journal={Phys. Rev.  B},
  volume={88},
  number={8},
  pages={085117},
  year={2013},
  publisher={APS}
}

@article{giannozzi2017advanced,
  title={Advanced capabilities for materials modelling with Quantum ESPRESSO},
  author={Giannozzi, Paolo and Andreussi, Oliviero and Brumme, Thomas and Bunau, Oana and Buongiorno Nardelli, M and Calandra, Matteo and Car, Roberto and Cavazzoni, Carlo and Ceresoli, Davide and Cococcioni, Matteo and others},
  journal={J. Phys. Condens. Matter.},
  volume={29},
  number={46},
  pages={465901},
  year={2017},
  publisher={IOP Publishing}
}

@article{giannozzi2009quantum,
  title={QUANTUM ESPRESSO: a modular and open-source software project for quantum simulations of materials},
  author={Giannozzi, Paolo and Baroni, Stefano and Bonini, Nicola and Calandra, Matteo and Car, Roberto and Cavazzoni, Carlo and Ceresoli, Davide and Chiarotti, Guido L and Cococcioni, Matteo and Dabo, Ismaila and others},
  journal={J. Phys. Condens. Matter.},
  volume={21},
  number={39},
  pages={395502},
  year={2009}
}

@article{gebauer2014pre,
  title={Pre-nucleation clusters as solute precursors in crystallisation},
  author={Gebauer, Denis and Kellermeier, Matthias and Gale, Julian D and Bergstr{\"o}m, Lennart and C{\"o}lfen, Helmut},
  journal={Chem. Soc. Rev.},
  volume={43},
  number={7},
  pages={2348--2371},
  year={2014},
  publisher={Royal Society of Chemistry}
}

@article{gebauer2008stable,
  title={Stable prenucleation calcium carbonate clusters},
  author={Gebauer, Denis and Volkel, Antje and Colfen, Helmut},
  journal={Science},
  volume={322},
  number={5909},
  pages={1819--1822},
  year={2008},
  publisher={American Association for the Advancement of Science}
}

@article{kimura2022possible,
  title={Possible embryos and precursors of crystalline nuclei of calcium carbonate observed by liquid-cell transmission electron microscopy},
  author={Kimura, Yuki and Katsuno, Hiroyasu and Yamazaki, Tomoya},
  journal={Faraday Discuss.},
  volume={235},
  pages={81--94},
  year={2022},
  publisher={Royal Society of Chemistry}
}

@article{piaggi2025ab,
  title={Ab initio machine-learning simulation of calcium carbonate from aqueous solutions to the solid state},
  author={Piaggi, Pablo M and Gale, Julian D and Raiteri, Paolo},
  journal={Proc. Natl. Acad. Sci.},
  volume={122},
  number={41},
  pages={e2415663122},
  year={2025},
  publisher={National Academy of Sciences}
}

@article{agmon2016protons,
  title={Protons and hydroxide ions in aqueous systems},
  author={Agmon, Noam and Bakker, Huib J and Campen, R Kramer and Henchman, Richard H and Pohl, Peter and Roke, Sylvie and Thamer, Martin and Hassanali, Ali},
  journal={Chemical reviews},
  volume={116},
  number={13},
  pages={7642--7672},
  year={2016},
  publisher={ACS Publications}
}

@article{marx2006proton,
  title={Proton transfer 200 years after von Grotthuss: Insights from ab initio simulations},
  author={Marx, Dominik},
  journal={ChemPhysChem},
  volume={7},
  number={9},
  pages={1848--1870},
  year={2006},
  publisher={Wiley Online Library}
}

@article{raiteri2015thermodynamically,
  title={Thermodynamically consistent force field for molecular dynamics simulations of alkaline-earth carbonates and their aqueous speciation},
  author={Raiteri, Paolo and Demichelis, Raffaella and Gale, Julian D},
  journal={J. Phys. Chem. C},
  volume={119},
  number={43},
  pages={24447--24458},
  year={2015},
  publisher={ACS Publications}
}

@article{raiteri2010derivation,
  title={Derivation of an accurate force-field for simulating the growth of calcium carbonate from aqueous solution: A new model for the calcite- water interface},
  author={Raiteri, Paolo and Gale, Julian D and Quigley, David and Rodger, P Mark},
  journal={J. Phys. Chem. C},
  volume={114},
  number={13},
  pages={5997--6010},
  year={2010},
  publisher={ACS Publications}
}

@article{car1985unified,
  title={Unified approach for molecular dynamics and density-functional theory},
  author={Car, Richard and Parrinello, Mark},
  journal={Phys. Rev. Lett.},
  volume={55},
  number={22},
  pages={2471},
  year={1985},
  publisher={APS},
  doi={10.1103/PhysRevLett.55.2471}
}

@article{demichelis2011stable,
  title={Stable prenucleation mineral clusters are liquid-like ionic polymers},
  author={Demichelis, Raffaella and Raiteri, Paolo and Gale, Julian D and Quigley, David and Gebauer, Denis},
  journal={Nat. Commun.},
  volume={2},
  number={1},
  pages={590},
  year={2011},
  publisher={Nature Publishing Group UK London}
}

@article{wallace2013microscopic,
  title={Microscopic evidence for liquid-liquid separation in supersaturated CaCO3 solutions},
  author={Wallace, Adam F and Hedges, Lester O and Fernandez-Martinez, Alejandro and Raiteri, Paolo and Gale, Julian D and Waychunas, Glenn A and Whitelam, Stephen and Banfield, Jillian F and De Yoreo, James J},
  journal={Science},
  volume={341},
  number={6148},
  pages={885--889},
  year={2013},
  publisher={American Association for the Advancement of Science}
}

@article{weiner2011crystallization,
  title={Crystallization pathways in biomineralization},
  author={Weiner, Steve and Addadi, Lia},
  journal={Annual review of materials research},
  volume={41},
  number={1},
  pages={21--40},
  year={2011},
  publisher={Annual Reviews}
}

@article{behler2007generalized,
  title={Generalized neural-network representation of high-dimensional potential-energy surfaces},
  author={Behler, J{\"o}rg and Parrinello, Michele},
  journal={Phys. Rev. Lett.},
  volume={98},
  number={14},
  pages={146401},
  year={2007},
  publisher={APS}
}

@article{tribello2014plumed,
  title={PLUMED 2: New feathers for an old bird},
  author={Tribello, Gareth A and Bonomi, Massimiliano and Branduardi, Davide and Camilloni, Carlo and Bussi, Giovanni},
  journal={Comput. Phys. Commun.},
  volume={185},
  number={2},
  pages={604--613},
  year={2014},
  publisher={Elsevier}
}

@article{plumed2019promoting,
  title={Promoting transparency and reproducibility in enhanced molecular simulations},
  journal={Nat. Methods},
  volume={16},
  number={8},
  pages={670--673},
  year={2019},
  publisher={Nature Publishing Group US New York}
}

@article{podryabinkin2017active,
  title={Active learning of linearly parametrized interatomic potentials},
  author={Podryabinkin, Evgeny V and Shapeev, Alexander V},
  journal={Comput. Mater. Sci.},
  volume={140},
  pages={171--180},
  year={2017},
  publisher={Elsevier}
}

@article{invernizzi2020unified,
  title={Unified approach to enhanced sampling},
  author={Invernizzi, Michele and Piaggi, Pablo M and Parrinello, Michele},
  journal={Phys. Rev.  X},
  volume={10},
  number={4},
  pages={041034},
  year={2020},
  publisher={APS}
}

@article{invernizzi2020rethinking,
  title={Rethinking metadynamics: from bias potentials to probability distributions},
  author={Invernizzi, Michele and Parrinello, Michele},
  journal={J. Phys. Chem. Lett.},
  volume={11},
  number={7},
  pages={2731--2736},
  year={2020},
  publisher={ACS Publications}
}

@article{ibrahim2024efficient,
  title={Efficient parametrization of transferable atomic cluster expansion for water},
  author={Ibrahim, Eslam and Lysogorskiy, Yury and Drautz, Ralf},
  journal={J. Chem. Theory Comput.},
  volume={20},
  number={24},
  pages={11049--11057},
  year={2024},
  publisher={ACS Publications}
}

@article{bochkarev2022efficient,
  title={Efficient parametrization of the atomic cluster expansion},
  author={Bochkarev, Anton and Lysogorskiy, Yury and Menon, Sarath and Qamar, Minaam and Mrovec, Matous and Drautz, Ralf},
  journal={Phys. Rev. Mater.},
  volume={6},
  number={1},
  pages={013804},
  year={2022},
  publisher={APS}
}

@article{drautz2019atomic,
  title={Atomic cluster expansion for accurate and transferable interatomic potentials},
  author={Drautz, Ralf},
  journal={Phys. Rev.  B},
  volume={99},
  number={1},
  pages={014104},
  year={2019},
  publisher={APS}
}

@article{lysogorskiy2021performant,
  title={Performant implementation of the atomic cluster expansion (PACE) and application to copper and silicon},
  author={Lysogorskiy, Yury and Oord, Cas van der and Bochkarev, Anton and Menon, Sarath and Rinaldi, Matteo and Hammerschmidt, Thomas and Mrovec, Matous and Thompson, Aidan and Cs{\'a}nyi, G{\'a}bor and Ortner, Christoph and others},
  journal={npj Comput. Mater.},
  volume={7},
  number={1},
  pages={97},
  year={2021},
  publisher={Nature Publishing Group UK London}
}

@article{grimme2010consistent,
  title={A consistent and accurate ab initio parametrization of density functional dispersion correction (DFT-D) for the 94 elements H-Pu},
  author={Grimme, Stefan and Antony, Jens and Ehrlich, Stephan and Krieg, Helge},
  journal={J. Chem. Phys.},
  volume={132},
  number={15},
  pages={154104},
  year={2010},
  publisher={American Institute of Physics}
}

@article{grimme2011effect,
  title={Effect of the damping function in dispersion corrected density functional theory},
  author={Grimme, Stefan and Ehrlich, Stephan and Goerigk, Lars},
  journal={J. Comput. Chem.},
  volume={32},
  number={7},
  pages={1456--1465},
  year={2011},
  publisher={Wiley Online Library}
}

@article{lysogorskiy2023active,
  title={Active learning strategies for atomic cluster expansion models},
  author={Lysogorskiy, Yury and Bochkarev, Anton and Mrovec, Matous and Drautz, Ralf},
  journal={Phys. Rev. Mater.},
  volume={7},
  number={4},
  pages={043801},
  year={2023},
  publisher={APS}
}

@article{perdew1996generalized,
  title={Generalized gradient approximation made simple},
  author={Perdew, John P and Burke, Kieron and Ernzerhof, Matthias},
  journal={Phys. Rev. Lett.},
  volume={77},
  number={18},
  pages={3865},
  year={1996},
  publisher={APS}
}

@article{sun2015strongly,
  title={Strongly constrained and appropriately normed semilocal density functional},
  author={Sun, Jianwei and Ruzsinszky, Adrienn and Perdew, John P},
  journal={Phys. Rev. Lett.},
  volume={115},
  number={3},
  pages={036402},
  year={2015},
  publisher={APS}
}

@article{LAMMPS,
  title={LAMMPS-a flexible simulation tool for particle-based materials modeling at the atomic, meso, and continuum scales},
  author={Thompson, Aidan P and Aktulga, H Metin and Berger, Richard and Bolintineanu, Dan S and Brown, W Michael and Crozier, Paul S and In't Veld, Pieter J and Kohlmeyer, Axel and Moore, Stan G and Nguyen, Trung Dac and others},
  journal={Comput. Phys. Commun.},
  volume={271},
  pages={108171},
  year={2022},
  publisher={Elsevier}
}

@book{kashchiev2000nucleation,
  title={Nucleation},
  author={Kashchiev, Dimo},
  year={2000},
  publisher={Elsevier}
}

@article{laio2002escaping,
  title={Escaping free-energy minima},
  author={Laio, Alessandro and Parrinello, Michele},
  journal={Proc. Natl. Acad. Sci.},
  volume={99},
  number={20},
  pages={12562--12566},
  year={2002},
  publisher={National Academy of Sciences}
}
\end{document}